\documentclass[11pt]{article}
\usepackage[dvips]{graphicx}
\begin{document}
\begin{center}
{\LARGE\bf  The Evolution of Universe with the B-I Type Phantom
Scalar Field }
 \vskip 0.15 in
$^\dag$Wei Fang, $^\ddag$H.Q.Lu, $^\ast$Z.G.Huang, K.F Zhang\\
Department~of~Physics,~Shanghai~University,\\
~Shanghai,~200444,~P.R.China\\
\footnotetext{$\dag$ Fangwei@graduate.shu.edu.cn}
\footnotetext{$\ddag$ Alberthq$_-$lu@staff.shu.edu.cn}
\footnotetext{$\ast$ Current address: Department of Mathematics
and Science,
   \\Huaihai Institute of Technology, Lianyungang,
222005, P.R.China }

 \vskip 0.5 in
\centerline{Abstract} \vskip 0.2 in
\begin{minipage}{5.5in}
{ \hspace*{15pt}\small We considered the phantom cosmology with a
lagrangian $\displaystyle L=\frac{1}{\eta}[1-\sqrt{1+\eta
g^{\mu\nu}\phi_{,~\mu}\phi_{,~\nu}}]-u(\phi)$, which is original
from the nonlinear Born-Infeld type scalar field with the
lagrangian $\displaystyle L=\frac{1}{\eta}[1-\sqrt{1-\eta
g^{\mu\nu}\phi_{,~\mu}\phi_{,~\nu}}]-u(\phi)$. This cosmological
model can explain the accelerated  expansion of the universe with
the equation of state parameter $w\leq-1$. We get a sufficient
condition for a arbitrary  potential to admit a late time
attractor solution: the value of potential $u(X_c)$ at the
critical point $(X_c,0)$ should be maximum and large than zero. We
study a specific potential with the form of
$u(\phi)=V_0(1+\frac{\phi}{\phi_0})e^{(-\frac{\phi}{\phi_0})}$ via
phase plane analysis and compute the cosmological evolution by
numerical analysis in detail. The result shows that the phantom
field survive till today (to account for the observed late time
accelerated
 expansion) without interfering with the nucleosynthesis of the
 standard model(the density parameter $\Omega_{\phi}\simeq10^{-12}$
 at the equipartition epoch), and also avoid the future collapse of
 the universe.

 }
\end{minipage}
\end{center}
\newpage
\section{Introduction}
 \hspace*{15pt}  A recent great finding in modern cosmology is that the universe is undergoing
  a accelerating expansion at present epoch
  and have a decelerated expansion phase in the recent past. This
  dynamics property is not only inferred in high redshift surveys
  of type Ia supernovae[1], but also independently implied from
  seven cosmic microwave background experiments(including the
  latest WMAP results)[2].  People have proposed many
  candidates for dark energy to fit the current
  observations, several important candidates  may be
  the cosmological constant (or vacuum density), quintessence[3](a time varying
  scalar field evolving in a specific potential) and the  "k-essence"(see literatures[4,5] and References).
  The major difference among these models is
  that they predict different equations of state parameter of dark energy and
  different accelerating style of present universe. Very recently, joint analysis of CMB+SN-Ia+HST+LSS data hold that the
    equation of state parameter can lie in the rang of  $-1.32<w<-0.82$
   with a best-fit value of $w\sim-1.04$ at the $2-\sigma$
 confidence levels, slightly preferring "phantom" models[6].
  Though phantom models[7] poses a challenge to the wildly
  accepted energy condition[8,9] and may lead to rapid vacuum decay[10],
  it is still very meaningful to study these models in
  the sense that it fits the
  observation very well, and also is phenomenologically interesting.
   \par Nonlinear Born-Infeld theory has been considered widely in string theory and cosmology.
    In 1934[11], Born and Infeld put forward a
theory of nonlinear electromagnetic field to resolve the
singularity in classical electromagnetic dynamics. The lagrangian
density is
\begin{equation}L_{BI}=b^2\left[1-\sqrt{1+(\frac{1}{2b^2})F_{\mu~\nu}F^{\mu~\nu}}~\right]\end{equation}
In order to describe the process of meson multiple production
connected with strong field regime[12], Heisenberg proposed the
following nonlinear scalar field lagrangian firstly:
\begin{equation}L=\frac{1}{\eta}\left[1-\sqrt{1-\eta
g^{\mu\nu}\phi_{,~\mu}\phi_{,~\nu}}~\right]\end{equation} This
lagrangian density(2) possesses some interesting characteristics:
(i)it is exceptional in the sense that shock waves do not develop
under smooth or continuous initial conditions[13], (ii)because
nonlinearities have been introduced, nonsingular scalar field
solutions can be generated, (iii) if
$g^{\mu\nu}\phi_{,~\mu}\phi_{,~\nu}<< \frac{1}{\eta}$, by Taylor
expansion, Eq.(2) approximates to the lagrangian of linear scalar
field,\begin{equation}\lim_{\eta\rightarrow
0}L=\frac{1}{2}g^{\mu\nu}\phi_{,~\mu}\phi_{,~\nu}
\end{equation} the linear theory is recovered. H.P.de Oliveira has
investigated qualitatively the static and spherically symmetric
solutions of this nonlinear scalar field[14]. One of our authors
has investigated the quantum cosmology based on the same nonlinear
B-I type scalar field[15]. Especially, if the potential $u(\phi)$
equals to $\frac{1}{\eta}$, we find that
$p\rho=-\frac{1}{\eta^2}$, this describes a  Chaplygin gas. The
Chaplygin gas has raised a renewed interest recently because of
its many remarkable and intriguing unique features[16].
 \par We have proposed a dark energy model based on the lagrangian Eq(2) in literature[8],
 and found the universe with this B-I type scalar field with a
 potential can undergo a phase of accelerating expansion, the
 corresponding equation of state parameter lies in the range of
 $-1<w<-\frac{1}{3}$,
 for the nonlinear B-I lagrangian with a negative
 kinetic energy term, when $u(\phi)$ is constant, the corresponding equation of state
 parameter can lie in the range $w<-1$.
In this paper, we consider the phantom cosmology of this nonlinear
B-I lagrangian, i.e, with a negative kinetic energy term,
$L=\frac{1}{\eta}\left[1-\sqrt{1+\eta
g^{\mu\nu}\phi_{,~\mu}\phi_{,~\nu}}~\right]-u(\phi)$, where
$u(\phi)$ is taken the form of
$u(\phi)=V_0(1+\frac{\phi}{\phi_0})e^{-\frac{\phi}{\phi_0}}$,
investigate the global structure of the
  dynamical system via phase plane analysis, and calculate the
  cosmological evolution by numerical analysis. The paper is organized as follows:
   In section 2,we
  start the phantom model with the B-I lagrangian, in section 3,
   the dynamical evolution of the phantom field without the presence of matter and
    radiation is considered, we discuss  the
  evolution of the phantom field at the presence of matter and
  radiation in section 4, section 5 is summary.

\section{The Phantom Model with B-I Lagrangian
}\hspace*{15pt}
  \par We consider the kinetic energy term is negative, namely the phantom
  cosmology.
  For the
spatially homogeneous phantom scalar field, we have the following
lagrangian:
\begin{equation}L=\frac{1}{\eta}\left[1-\sqrt{1+\eta\dot{\phi}^2}~\right]-u(\phi)\end{equation}
where $u{(\phi)}$ is the positive potential, $\eta$ is a constant.
First we consider the simple case where the phantom field $\phi$
is dominant only. In the spatially flat Robertson-Walker metric
$ds^2=dt^2-a^2(t)(dx^2+dy^2+dz^2)$, the Einstein equation
$G_{\mu\nu}={\xi}T_{\mu\nu}$, can be written as
 \begin{equation}\left\{\begin{array}{l}
 H^2=\frac{\xi}{3}\rho_{\phi}\\
        \frac{\ddot{a}}{a}=-\frac{\xi}{6}(\rho_{\phi}+3P_{\phi})
                                                  \end{array}
                                                  \right.
                                                  \end{equation}
        For a spatially homogenous phantom field $\phi$, we have
        the equation of motion
        \begin{equation}\ddot{\phi}+3H\dot{\phi}(1+\eta\dot{\phi}^2)-u^{'}(\phi)(1+\eta\dot{\phi}^2)^\frac{3}{2}=0\end{equation}
        where the overdot represents the differentiation with
        respect to t and the prime denotes the differentiation
        with respect to $\phi$, the constant $\xi=8{\pi}G$, where
        $G$ is a Newtonian gravitation constant.
 The Energy-moment tensor is
\begin{equation}\displaystyle T^\mu_\nu=-\frac{g^{\mu\rho}
\phi_{,~\nu}\phi_{,~\rho}}{\sqrt{1+\eta
g^{\mu\nu}\phi_{,~\mu}\phi_{,~\nu}}}-\delta^\mu_\nu
L\end{equation} From Eq.(7), we have
\begin{equation}\rho_\phi=T^0_0=\frac{1}{\eta\sqrt{1+\eta
\dot{\phi}^2}}-\frac{1}{\eta}+u\end{equation}
\begin{equation}P_\phi=-T^i_i=\frac{1}{\eta}-\frac{\sqrt{1+\eta
\dot{\phi}^2}}{\eta}-u\end{equation} From Eqs.(8) and (9), we
obtain
\begin{equation}\rho_\phi+p_\phi=-\frac{\dot{\phi}^2}{\sqrt{1+\eta
\dot{\phi}^2}}\end{equation} It is clear that the equation of
state parameter $w_\phi<-1$ is completely confirmed by Eq.(10)
which agrees with the recent analysis of observation data. We also
can get
\begin{equation}\rho_\phi+3p_\phi=\frac{2}{\eta}-\frac{2}{\eta}\sqrt{1+\eta\dot\phi^2}-\frac{\dot\phi^2}{\sqrt{1+\eta\dot\phi^2}}-2u\end{equation} It is obvious that
$\rho_\phi+3p_\phi<0$. Eq.(11)shows that the universe is
undergoing a phase of accelerating expansion. The model of phantom
B-I scalar field without potential $u(\phi)$ is hard to
understand. We can always find
$\displaystyle\rho_\phi=\frac{1}{\eta\sqrt{1+\eta\dot{\phi}^2}}-\frac{1}{\eta}<0$
in this case, it is unreasonable apparently. However, in the model
of phantom B-I scalar field with potential $u(\phi)$, when
 \begin{equation}
u(\phi)>\frac{1}{\eta}-\frac{1}{\eta\sqrt{1+\eta\dot{\phi}^2}}\end{equation}
$\rho_\phi$ is always greater than zero. In phantom B-I scalar
model with a potential $u(\phi)$, we also find the strong and weak
energy condition always failed from Eqs.(10)and (11). The equation
of state is
\begin{equation}w_\phi=\frac{P_\phi}{\rho_\phi}=-1-\frac{\eta\dot{\phi}^2}{1+[\eta
u(\phi)-1]\sqrt{1+\eta\dot{\phi}^2}}\end{equation} It is always
less than $-1$ if Eq(12) is satisfied.
\par We also can write the speed of sound in these phantom model:
\begin{equation}c^2_s=\frac{P_{,~X}}{\rho_{,~X}}=\frac{L_{\phi~,~X}}{L_{\phi~,~X}+2X{L_{\phi~,~X~X}}}\end{equation}
where $P=L_\phi(\phi,X)$ and $\rho=2XL_{\phi,X}-L_\phi(\phi,X)$
with $X=\frac{1}{2}(\partial_\mu\phi)^2$. In our model
$X=\frac{1}{2}\dot{\phi}^2$,  so the speed of sound of the phantom
with BI lagrangian is
\begin{equation}c^2_s=\frac{P_{,~X}}{\rho_{,~X}}=1+\eta\dot{\phi}^2\end{equation}
It is clear from Eq.(15) that the $c_s$ is always greater than $1$
unless $\dot\phi^2=0$, this means that the perturbation of the
background field can travel faster than light as measured in the
preferred frame where the background field is homogeneous. For a
time dependent background field, this is not a Lorentz invariant
state. However, it does not violate causality because the
underlying theory is manifestly Lorentz invariant and it is not
possible to transmit information faster than light along arbitrary
spacelike directions or create closed timelike curves[17]. In
Ref[18], they also shows that a fluid with $\mid w\mid>1$ does not
constradict causality if $w$ is not constant. In our model, $w$ is
clear a time-dependent parameter.
\section{Dynamical Evolution Of The Phantom Field}
\hspace*{15pt} As we consider the phantom field becomes dominant
,we can neglect the nonrelativitic and relativistic components
(matter and radiation) in the universe, then from Eq.(5,8,6), we
have
\begin{equation}\ddot{\phi}+\dot{\phi}(1+\eta\dot{\phi}^2)\sqrt{3\xi[\frac{1}{\eta\sqrt{1+\eta\dot{\phi}^2}}-\frac{1}{\eta}+u(\phi)]}-u^{'}(\phi)(1+\eta\dot{\phi}^2)^\frac{3}{2}=0\end{equation}
to study an numerical computation, it is convenient to introduce
two independent variables
\begin{equation}\left\{\begin{array}{l}
                                                 X=\phi\\
                                                 Y=\dot{\phi}
                                                  \end{array}
                                                  \right.
                                                  \end{equation} then Eq.(16) can be
written\begin{equation}\left\{\begin{array}{l}\frac{dX}{dt}=Y \\
                                                  \frac{dY}{dt}=u^{'}(X)(1+\eta Y^2)^\frac{3}{2}-Y (1+\eta Y^2)\sqrt{3\xi[\frac{1}{\eta\sqrt{1+\eta
                                                  Y^2}}-\frac{1}{\eta}+u(X)]}
                                                  \end{array}
                                                  \right.
                                                  \end{equation}
   we can obtain this system's critical point from\begin{equation}\left\{\begin{array}{l}\frac{dX}{dt}=0 \\
                                                  \frac{dY}{dt}=0
                                                  \end{array}
                                                  \right.
                                                  \end{equation}
   then its critical point is $(X_c,0)$, where the critical value $X_c$ is determined by $u'(X_c)=0$.
   Linearizing  Eq.(18) around the critical point, we have\begin{equation} \left\{\begin{array}{l}\frac{dX}{dt}=Y \\
                                                  \frac{dY}{dt}=u''(X_c)(X-X_c)-\sqrt{(3\xi u(X_c)}Y
                                                  \end{array}
                                                  \right.
                                                  \end{equation}
 the types of the critical point are determined by the eigenequation  of
 system\begin{equation}\lambda^2+\alpha\lambda+\beta=0\end{equation}
where $\alpha=\sqrt{3\xi u(X_c)}$,$\beta=-u''(X_c)$, the two
eigenvalues are $\lambda_1=\frac{-\sqrt{3\xi u(X_c)}+\sqrt{3\xi
u(X_c)+4u''(X_c)}}{2}$, $\lambda_2=\frac{-\sqrt{3\xi
u(X_c)}-\sqrt{3\xi u(X_c)+4u''(X_c)}}{2}$. For a positive
potentials, if $u''(X_c)<0$, then the critical point ($X_c$,0) is
a stable node, which implies that the dynamical system admits
attractor solutions. We can also conclude that if a potential
possesses the general properties: $u(X_c)>0,u'(X_c)=0$ and $
u''(X_c)<0$, then our phantom model with this potential must have
a attractor solution and will predict a late time de-sitter like
behavior($w_\phi=-1$). Some authors have studied the cosmological
dynamics of phantom scalar field with special potentials which
possessed above properties, such as
$u(\phi)=V_0e^{-\phi^2/\sigma^2}$[19],
$u(\phi)=V_0[\cosh(\frac{\alpha\phi}{M_p}]^{-1}$[20],
$u(\phi)=V_0[1+\alpha\phi^2/M_p]^{-1}$.
 \par Next we specify the potential with a special
 form. We choose a widely studied potential[21] as
 \begin{equation}u(\phi)=V_0(1+\frac{\phi}{\phi_0})e^{(-\frac{\phi}{\phi_0})}\end{equation}
 where $V_0>0$.
  It is easy to find that the critical $X_c=0$ and
 $u''(X_c)=-\frac{V_0}{\phi^2_0}<0$ in such a potential. Therefore
 this model has an attractor solution which corresponds to its
 attractor regime, the equation of state $w\leq-1$.
 Substitute Eq.(22) into Eq.(18), we obtain\begin{equation}\left\{\begin{array}{l}\frac{dX}{dt}=Y \\
                                                  \frac{dY}{dt}=-\frac{V_0}{\phi^2_0}X e^{(-\frac{X}{\phi_0})}(1+\eta
                                                  Y^2)^\frac{3}{2}
                                                  -Y (1+\eta Y^2)\sqrt{3\xi[\frac{1}{\eta\sqrt{1+\eta
                                                  Y^2}}-\frac{1}{\eta}+V_0(1+\frac{X}{\phi_0})e^{(\frac{X}{-\phi_0})}]}
                                                  \end{array}
                                                  \right.
                                                  \end{equation}
 We will solve this equations system via the numerical approach, to do this,
 we rescale the quantities as $x=\frac{X}{\phi_0}$,
 $s=(\phi^2_0\eta)^{-\frac{1}{2}}t$, $y=\sqrt{\eta }Y$,
 Then Eq.(22) becomes \begin{equation}\left\{\begin{array}{l}\frac{dx}{ds}=y \\
                                                  \frac{dy}{ds}=-\gamma x e^{(-x)}(1+
                                                  y^2)^\frac{3}{2}
                                                  -\phi_0 y(1+y^2)\sqrt{3\xi[\frac{1}{\sqrt{1+
                                                  y^2}}-1+\gamma(1+x)e^{(-x)}]}
                                                  \end{array}
                                                  \right.
                                                  \end{equation}
 where $\gamma=V_0\eta$
 are parameters, we take $\xi=1$ for convenience. The numerical results with different initial condition are plotted in
 Figs.$1-3$ and the parameters $\phi_0=\sqrt{0.1}, \gamma=3$.
\begin{center}\vspace{0.5cm}
\includegraphics[angle=270,width=10cm]{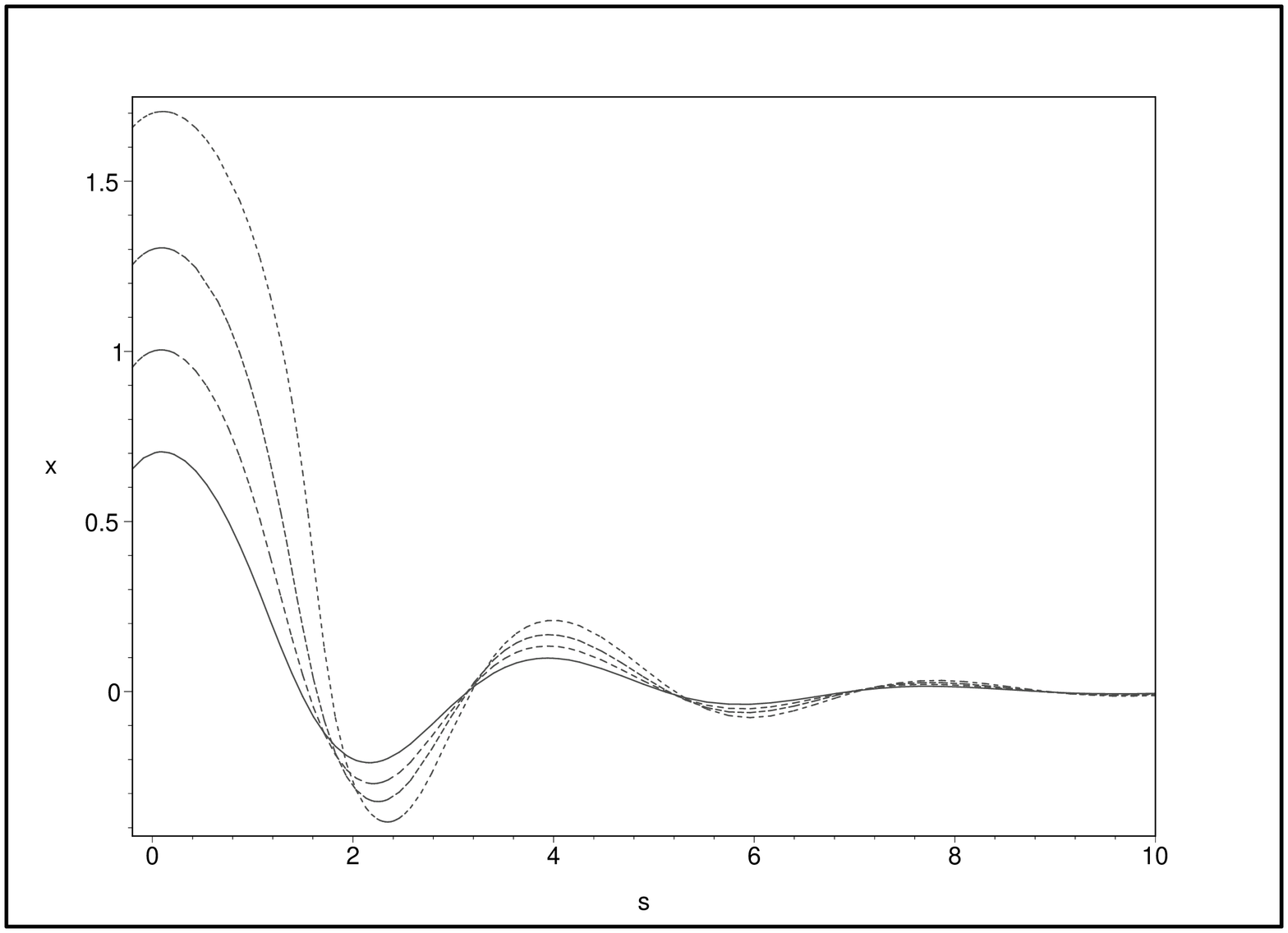} \end{center} \hfill
~\begin{minipage}{5.5in}~Fig1.This plot shows the evolution of the
scalar field in difference initial condition, solid line is for
$\phi_{in}=0.7\phi_0$, dotted line is for $\phi_{in}=\phi_0$,
dashed line and dot-dashed line for$\phi_{in}=1.3\phi_0,
1.7\phi_0$ respectively, they are all plotted for a fixed value of
$y_{in}=0.1$ .\end {minipage}
\begin{center}\vspace{0.5cm}
\includegraphics[angle=270,width=10cm]{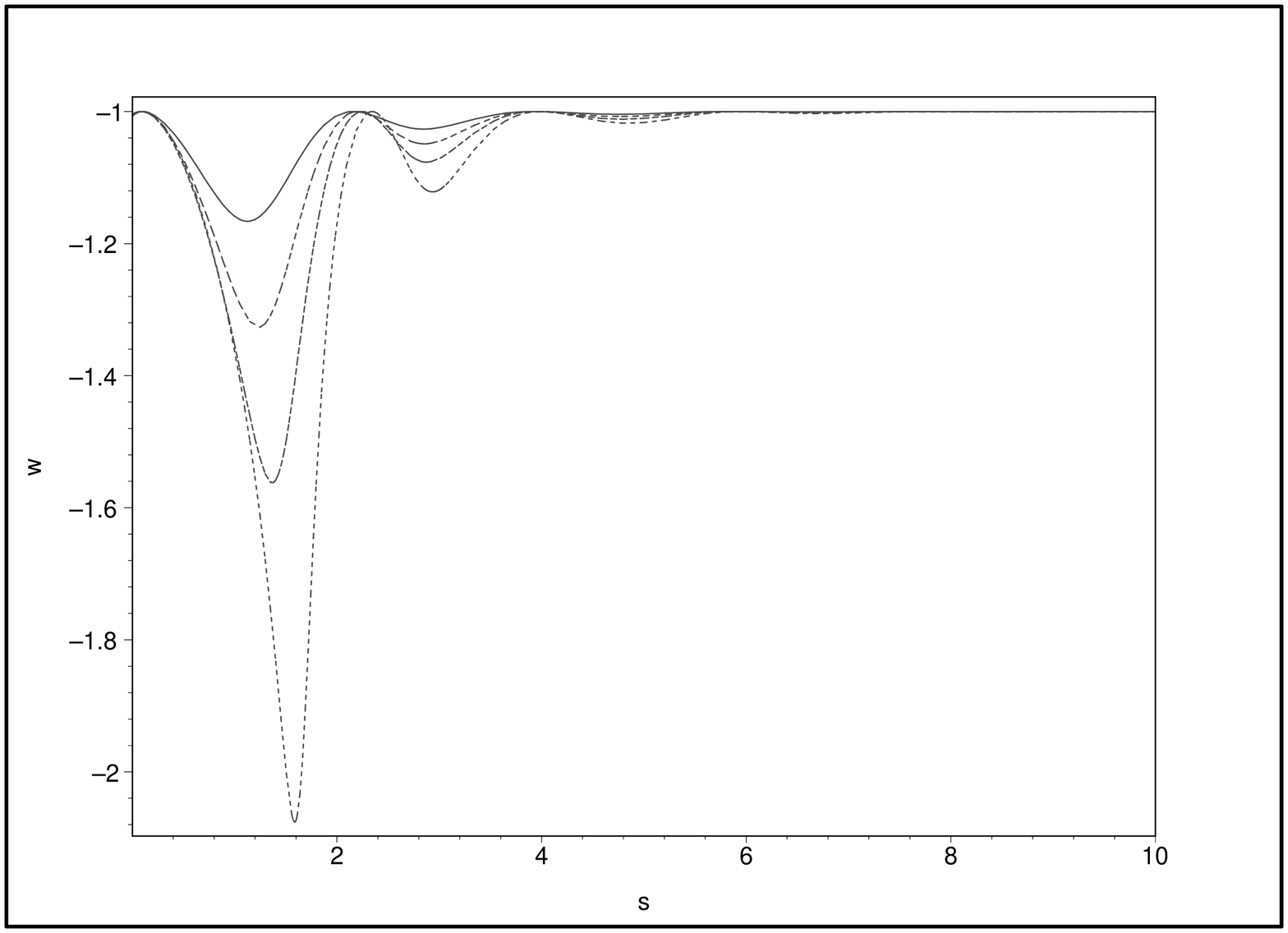} \end{center} \hfill
~\begin{minipage}{5.5in}~Fig2. The evolution of $w$ with respect
to s, the initial condition is the same as fig1.\end {minipage}
\begin{center}\vspace{0.5cm}
\includegraphics[angle=270,width=10cm]{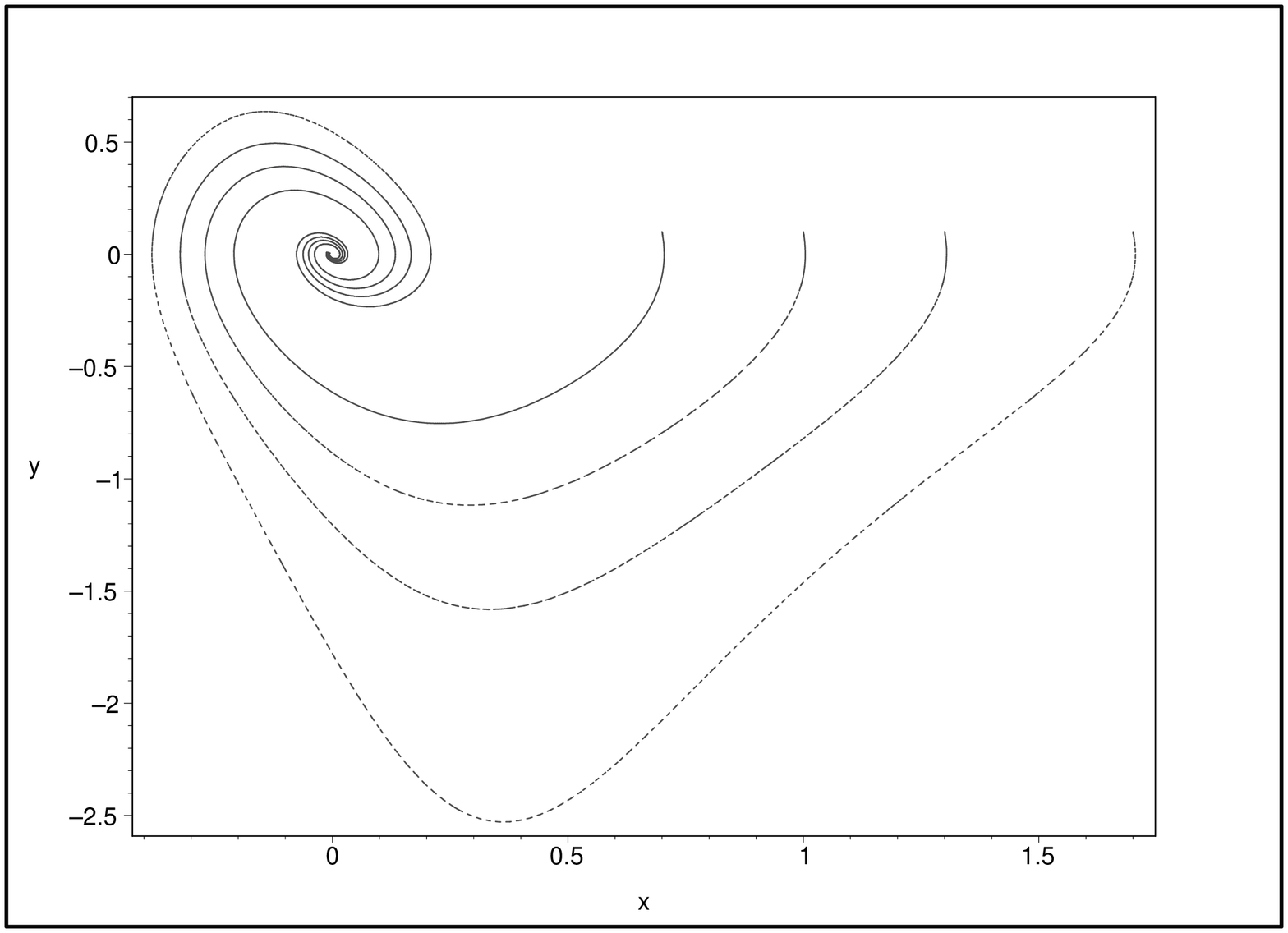} \end{center} \hfill
\begin{minipage}{5.5in}~Fig3. The attractor property of the
system in the phase plane, the initial condition is the same as
fig1.
\end{minipage}
\\
\par As we know,when $\eta\rightarrow0$, our model come back quintessence case. In order to see the nonlinear effect,  we  plot the
quintessence model with our model in fig4 and fig5.
 \begin{center}\vspace{0.5cm}
\includegraphics[angle=270,width=10cm]{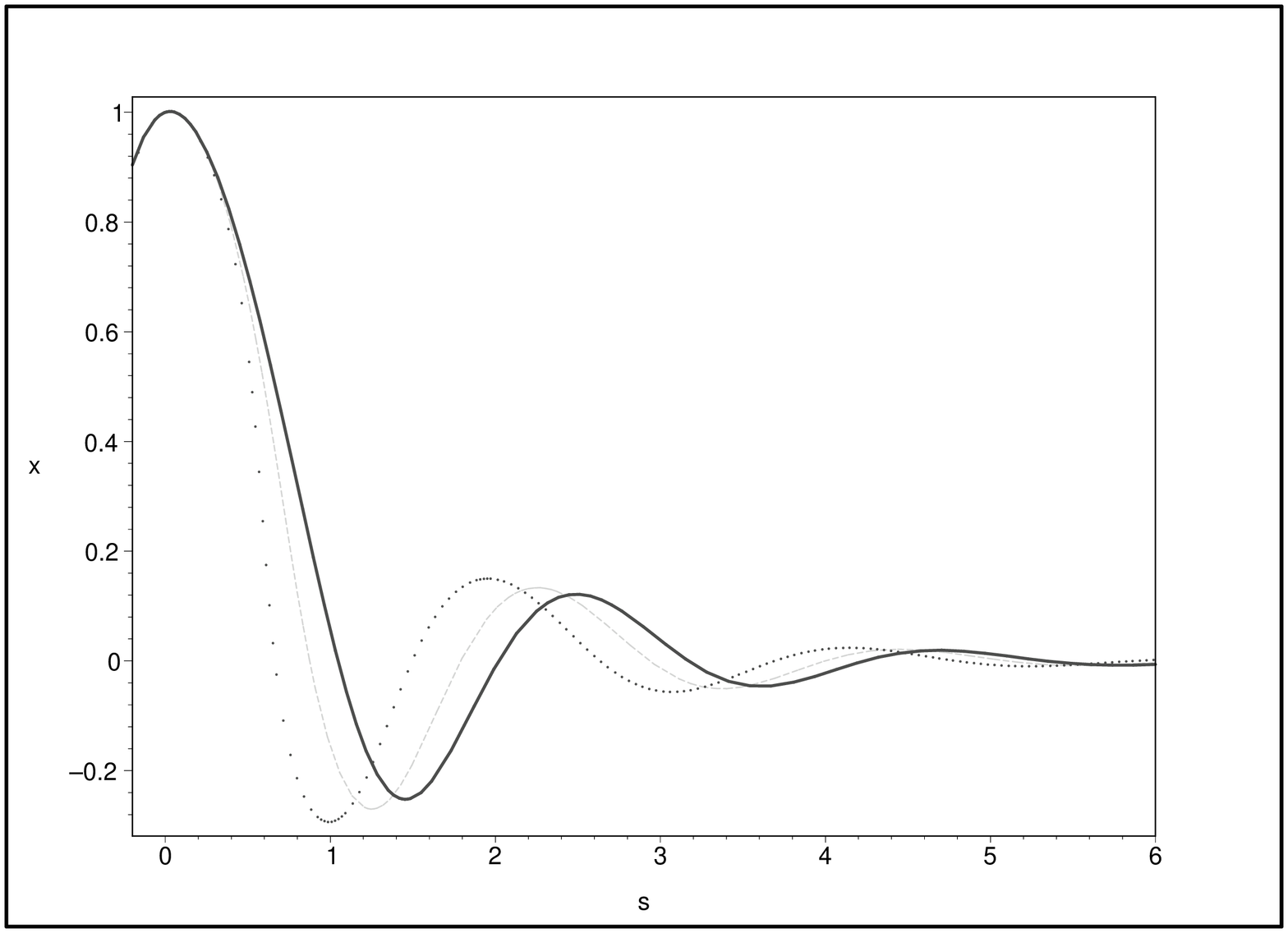} \end{center} \hfill
\par\begin{minipage}{5.5in} Fig4. the evolution of scalar field with respect to s, solid line is
quintessence model, dotted line and dashed line are nonlinear
scalar field,dotted line is for $\eta=2/3$,dashed line is for
$\eta=1/3$.
\end{minipage}
\begin{center}\vspace{0.5cm}
\includegraphics[angle=270,width=10cm]{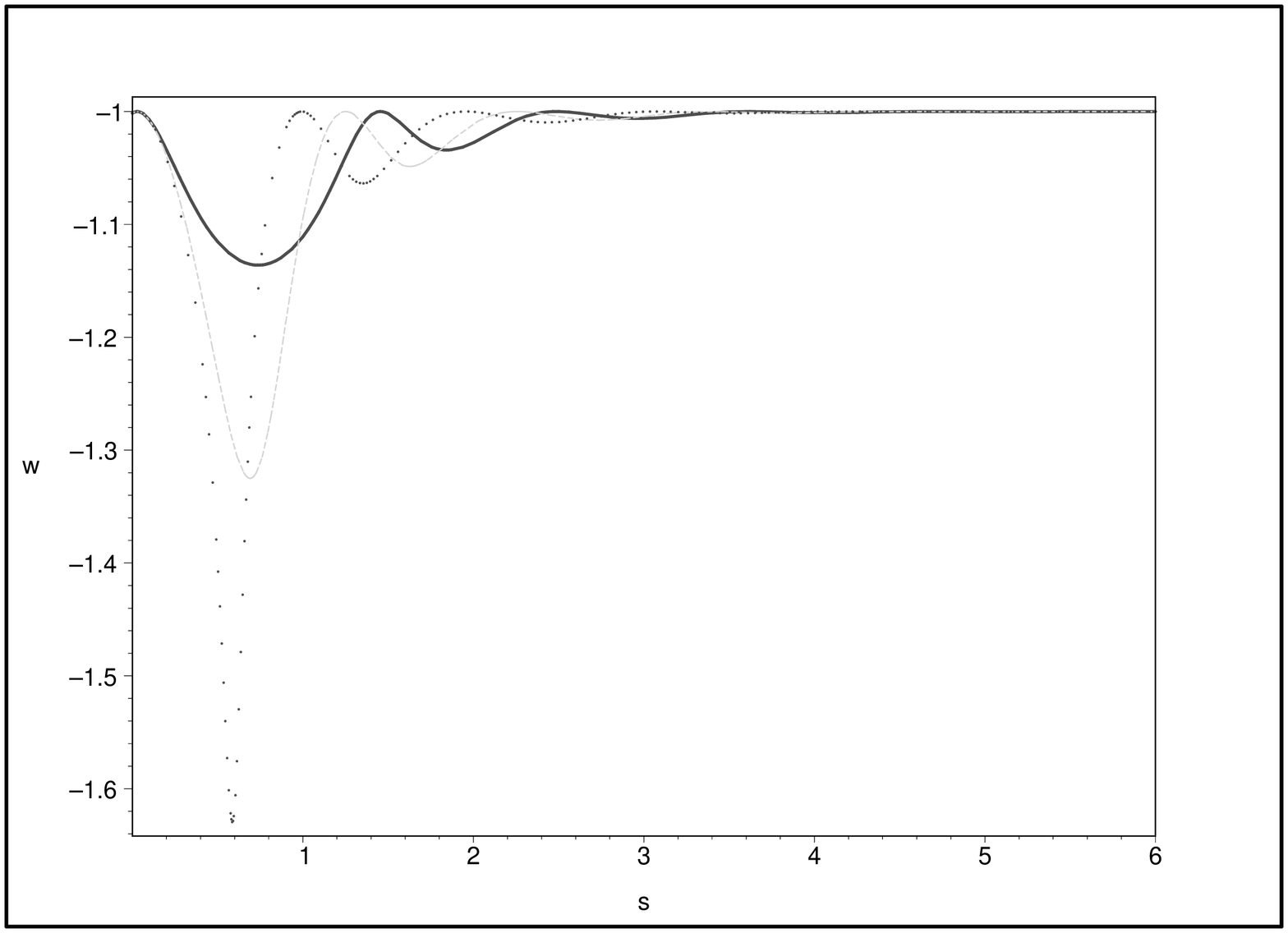} \end{center} \hfill
~\begin{minipage}{5.5in} Fig5.the evolution of $w$ with respect to
s, solid line is quintessence model, dotted line and dashed line
are nonlinear scalar field, dotted line is for $\eta=2/3$, dashed
line is for $\eta=1/3$.\end{minipage}
\\
\par From fig1,fig2 and fig3, we can easily find the system admit
a attractor solution, the equation of state parameter $w$ start
from nearly -1, then quickly evolve to the regime of smaller than
-1, turn back to execute the damped oscillation, and reach to -1
 eventually for ever. Due to the unusual physical properties in
phantom model, the phantom field, released at a distance from the
origin with a small kinetic energy, moves towards the top of the
potential and crosses over to the other side and then turns back
to execute the damped oscillation around the critical point, after
a certain time the motion ceases and the phantom field settles on
the top of the potential permanently to mimic the de Sitter-like
behavior($w_{\phi}=-1$). Fig4 and fig5 indeed shows that the
nonlinear scalar field will come back to quintessence case when
$\eta$ decreases to zero. the nonlinear effect does not affect the
global attractor behavior but change the evolution of $w$ and
scalar field $\phi$ in details.

 \section{Evolution of the phantom at the presence of matter and
 radiation}
 \hspace*{15 pt} In the previous section, we studied the case when
 phantom dominates over all other energy density.
Now we consider the more general case that the effect of matter
and radiation can not be neglected, thus the Eq.(5)becomes
\begin{equation}
 H^2=\frac{\xi}{3}{(\rho_{\phi}+\rho_M+\rho_r)}
                                                  \end{equation}
 where $\rho_M$ and $\rho_r$ denote the energy
 density of nonrelativistic and
 relativistic components(namely, matter and radiation), respectively. We know
 that\begin{equation}\left\{\begin{array}{l}\rho_Ma^3=\rho_{Mi}a^3_i=constant\\
                                                \rho_ra^4=\rho_{ri}a^4_i=constant
                                                  \end{array}
                                                  \right.
                                                  \end{equation}  So we can rewrite Eq.(25) as \begin{equation}
 H^2=H^2_i[\frac{\rho_\phi}{\rho_{ci}}
 +\Omega_{Mi}(\frac{a_i}{a})^3+\Omega_{ri}(\frac{a_i}{a})^4]
                                                  \end{equation}
 where $H^2_i=\frac{\xi\rho_{ci}}{3}$, $\rho_{ci}$ is the critical
 energy density
 of the universe at the initial $t_i$, $\rho_{mi}, \rho_{ri}$ are the
 matter and the radiation energy densities at $t_i$,  $\Omega_{mi},
 \Omega_{ri}$ are the cosmic density parameters for matter and
 radiation at $t_i$, $a_i$ is the initial scale factor at $t_i$, we
 will specify $a_i=1$ for convenience.
 We can finally give the form of Eq.(27) as\begin{equation}
 H^2=H^2_i[\frac{\rho_\phi}{\rho_{ci}}
 +\Omega_{Mi}a^{-3}+\Omega_{ri}a^{-4}]
                                                  \end{equation}
 we introduce the new
 variables $x=\frac{\phi}{\phi_0}, y=\sqrt\eta\frac{d\phi}{dt}, N=lna$, we
 can express Eq.(16) as\begin{equation}\left\{\begin{array}{l}\frac{dx}{dN}=\frac{y}{H_i\phi_0\eta}\psi \\
                                                  \frac{dy}{dN}=-3y(1+
                                                  y^2)+\sqrt\eta\frac{u'(\phi)(1+
                                                  y^2)^{\frac{3}{2}}}{H_i}\psi
                                                  \end{array}
                                                  \right.
                                                  \end{equation}
 where $\psi(x,y,N)=[\frac{1}{\rho_{ci}}(\frac{1}{\eta\sqrt{1+\eta \dot{\phi}^2}}-\frac{1}{\eta}+u)+\Omega_{Mi}e^{-3N}+\Omega_{ri}e^{-4N}]^{-\frac{1}{2}}$.
  Substitute the potential Eq.(22) into Eq.(29), we get\begin{equation}\left\{\begin{array}{l}\frac{dx}{dN}=\frac{y}{H_i\phi_0\sqrt\eta}\psi' \\
                                                  \frac{dy}{dN}=-3y(1+
                                                  y^2)-\frac{\sqrt\eta V_0}{\phi_0H_i}x(1+
                                                  y^2)^{\frac{3}{2}}e^{-x}\psi'
                                                  \end{array}
                                                  \right.
                                                  \end{equation}
  where $\psi'(x,y,N)=[\frac{V_0 \xi}{3H^2_i}(\frac{1}{\eta V_0\sqrt{1+y^2}}-\frac{1}{\eta
  V_0}+(1+x)e^{(-x)})+\Omega_{Mi}e^{-3N}+\Omega_{Mi}e^{-4N}]^{-\frac{1}{2}}$.
   The numerical results are plotted in Figs.$6-10$. We specify
   our starting point as the equipartition epoch, at
   which $\Omega_{Mi}=\Omega_{ri}=0.5$, We plot the results by
   choosing the parameter $\frac{V_0\xi}{H^2_i}=10^{-8}$.
\begin{center} \vspace{0.5cm}
\includegraphics[angle=270,width=10cm]{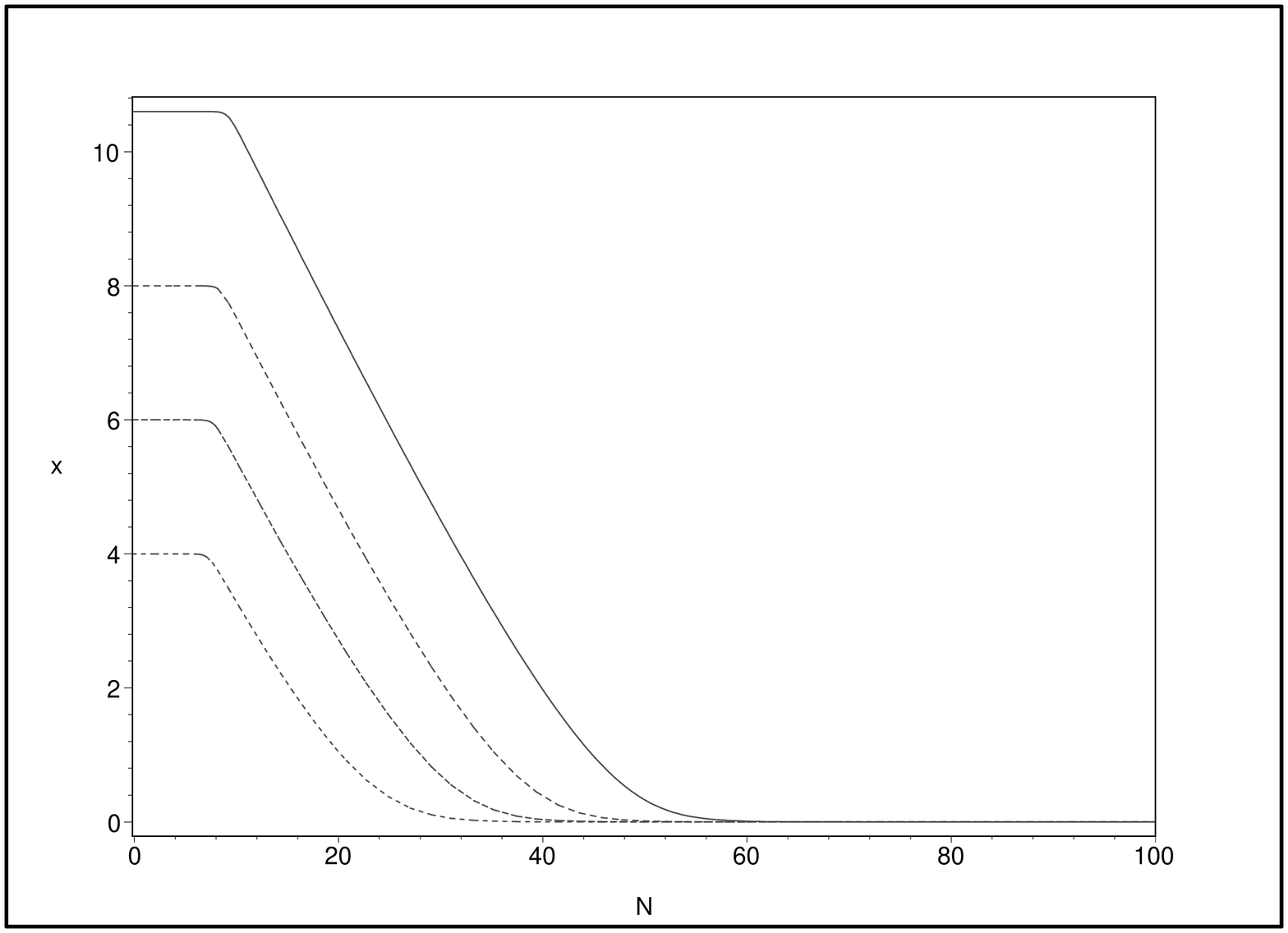}\end{center} \hfill
\begin{minipage}{5.5in} Fig6. The evolution of the scalar field  with
respect to N in the presence of matter and radiation. they are
plotted in different initial conditions, solid line is for
$\phi_{in}=10.6\phi_0$,dotted line ,dashed line and dot-dashed
line for $\phi_{in}=8\phi_0, 6\phi_0, 4\phi_0$ respectively,
$y_{in}$ is set zero all the time.
\end {minipage}
\begin{center}\vspace{0.5cm}
\includegraphics[angle=270,width=10cm]{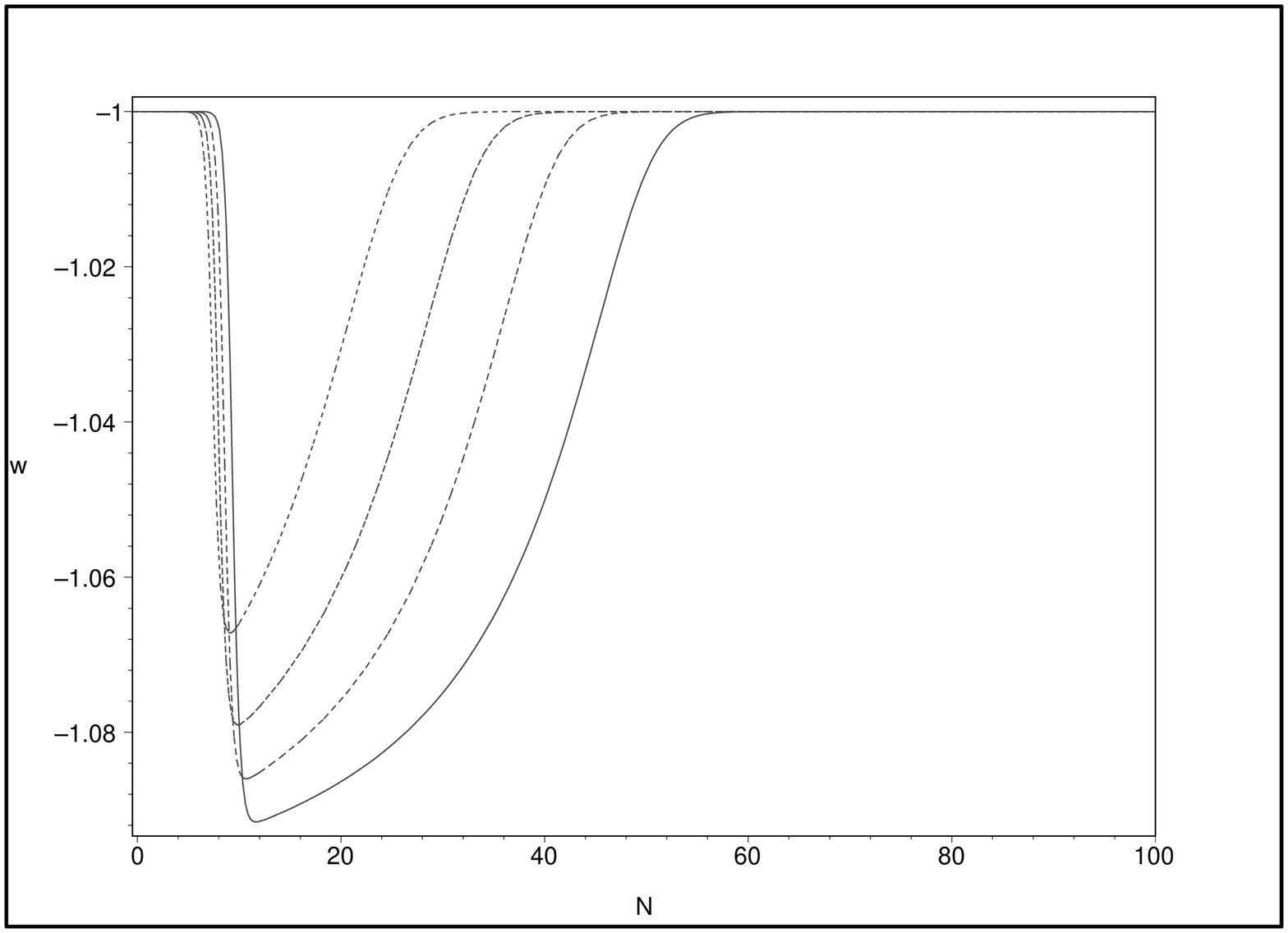}\end{center} \hfill
\begin{minipage}{5.5in} Fig7. The evolution of $w$ with respect to N in the presence of matter and
radiation, we chose the same initial condition as fig6.\end
{minipage}
 \begin{center}\vspace{0.5cm}
\includegraphics[angle=270,width=10cm]{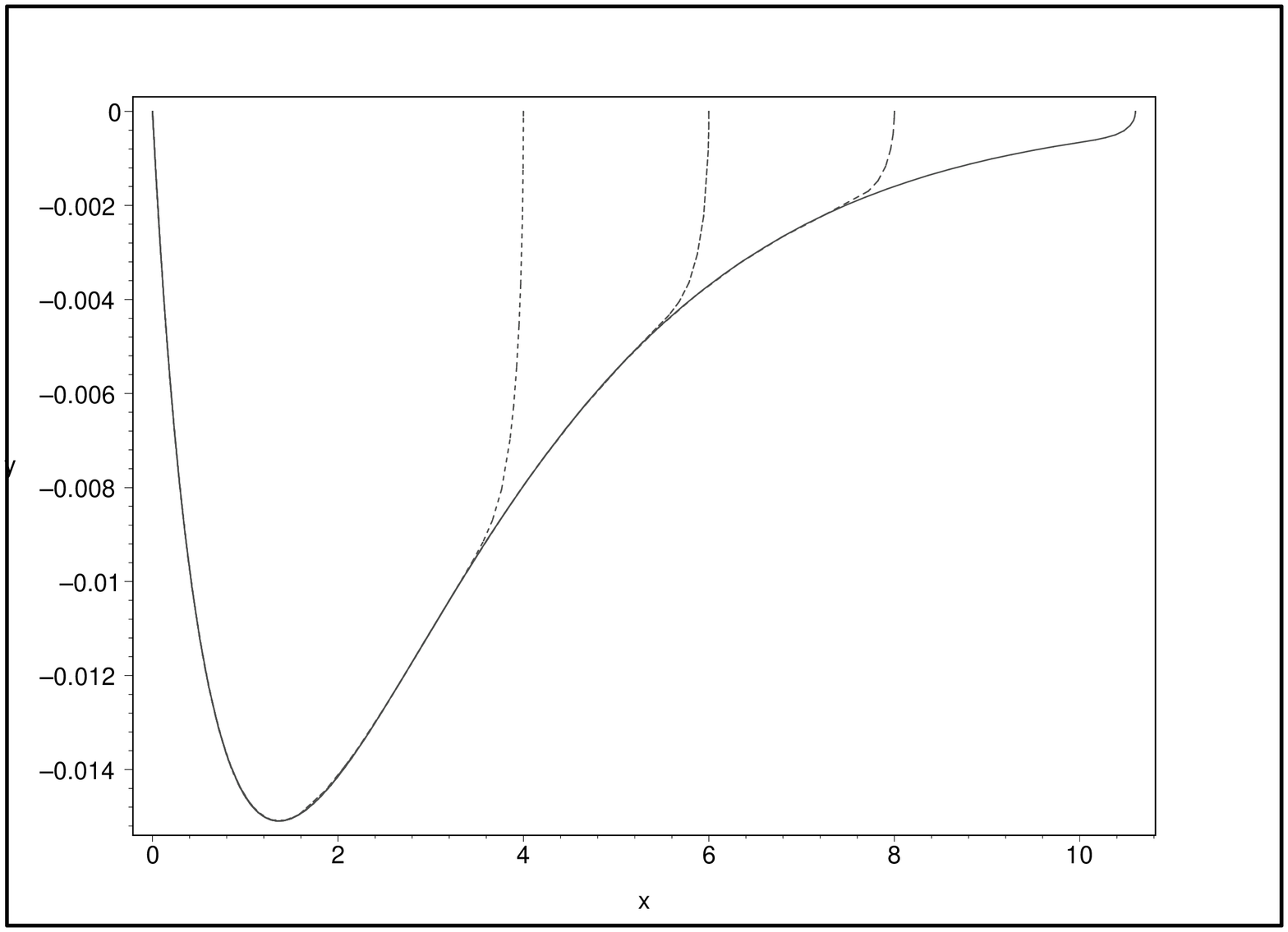}\end{center} \hfill
\begin{minipage}{5.5in} Fig8. The attractor property of the scalar field in the
presence of matter and radiation, the initial condition is the
same as in fig6.\end {minipage}
 \begin{center}\vspace{0.5cm}
\includegraphics[angle=270,width=10cm]{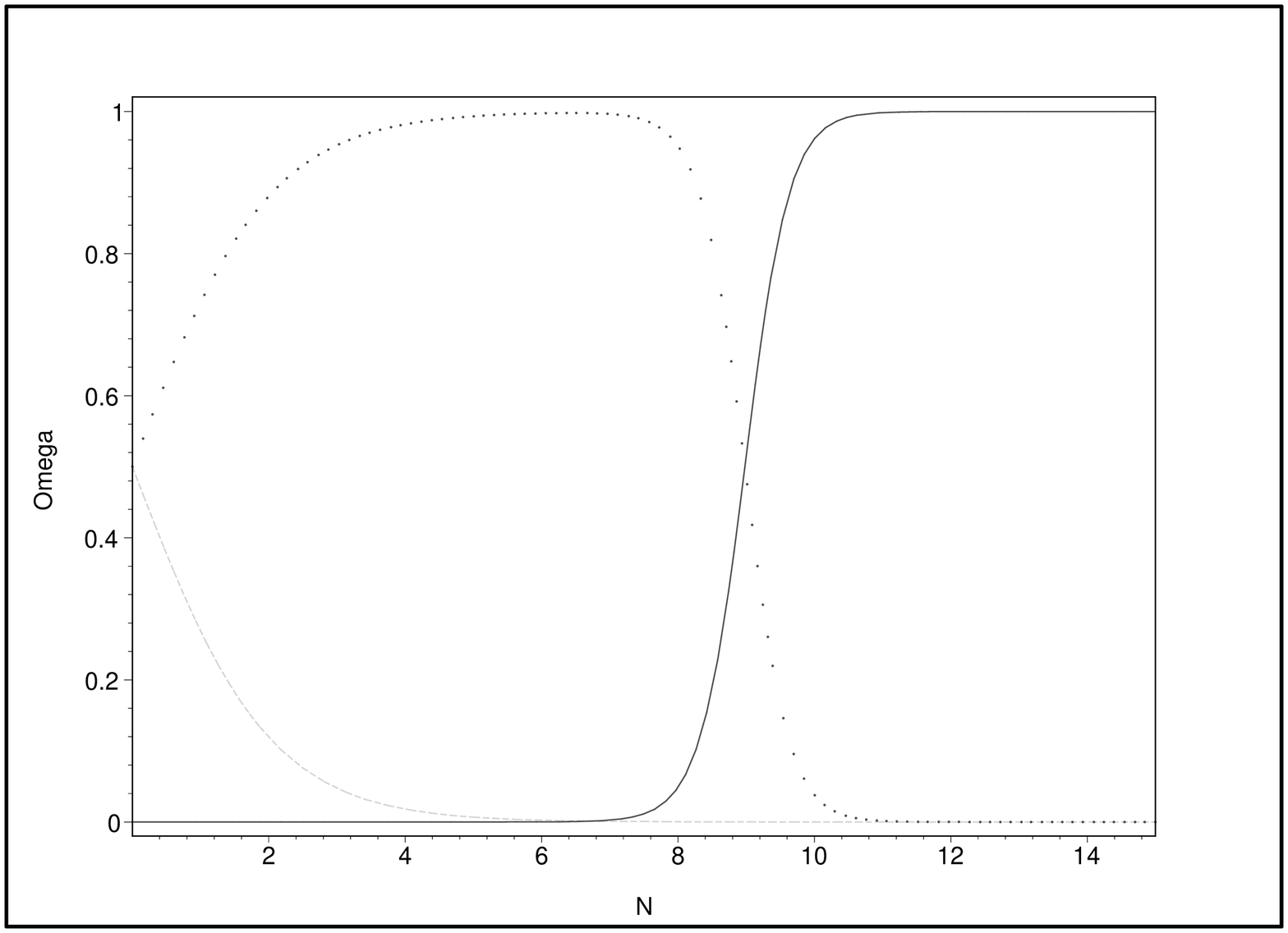} \end{center}\hfill
\begin{minipage}{5.5in} Fig9. The evolution of density parameter $\Omega$ with respect to N
, solid line is for scalar field ,dotted line for matter and
dashed line for radiation, where the initial value of
$\phi_{in}=10.6\phi_0, y_{in}=0$.
\end {minipage}
 \begin{center}\vspace{0.5cm}
\includegraphics[angle=270,width=10cm]{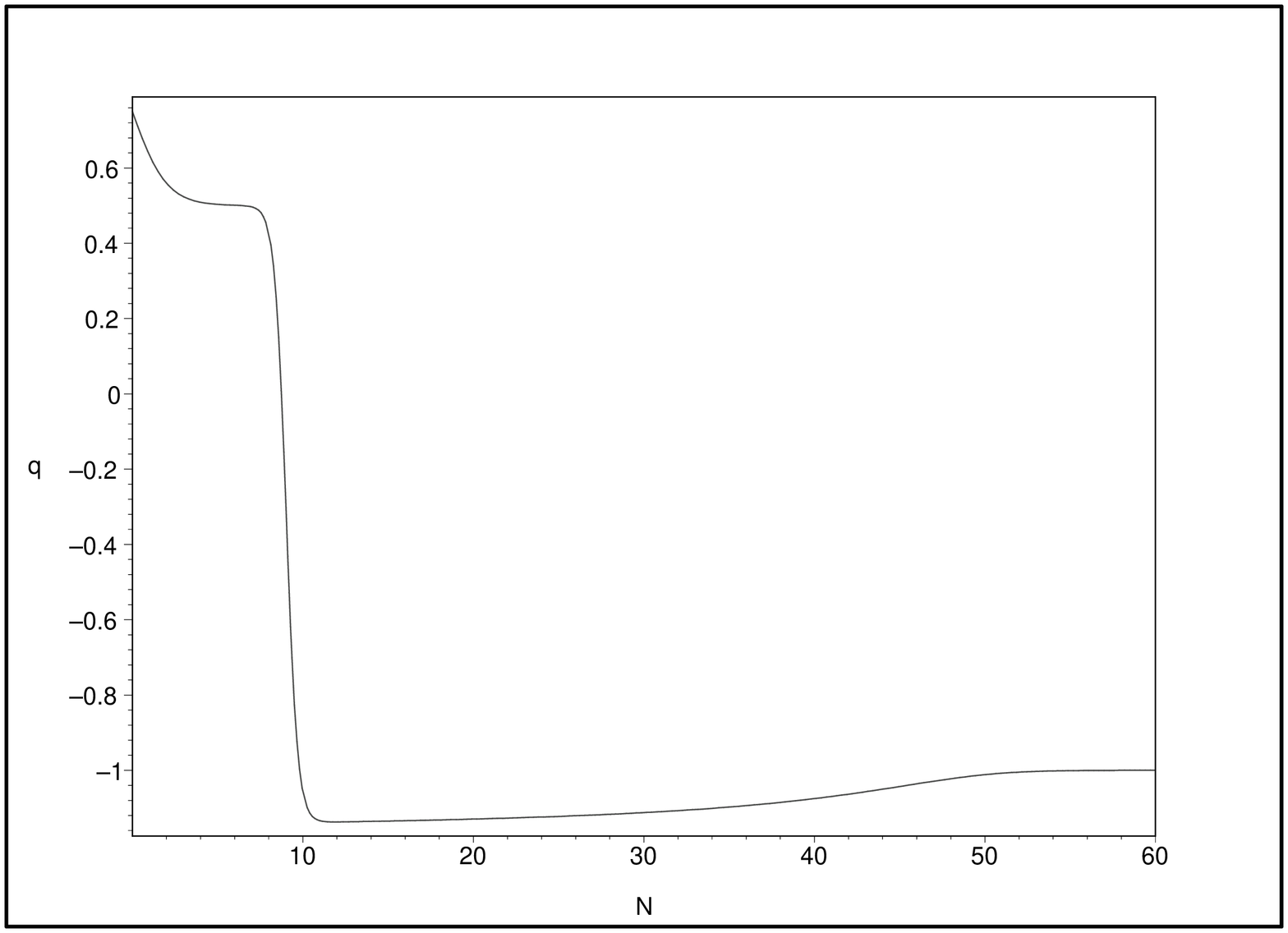} \end{center}\hfill
\begin{minipage}{5.5in} Fig10. The evolution of decelerating factor $q$ with respect to N, the initial condition
is the same as fig9. \end {minipage}
\par From these results(fig6,7,8) we can
 conclude that there is still a late time attractor solution even
 in the presence of matter and radiation.
  The phantom field energy density will dominant over the matter and radiation
  composition($\Omega_{\phi}\rightarrow1$)with its state parameter
  $w_{\phi}\rightarrow-1$, mimicking the cosmological
  constant. Such evolution behaviors will avoid
  the cosmic doomsday [9].  Fig10 quantitatively shows the universe is speeding
  up at present but undergoes a decelerated regime at the recent
  past. Since $T_{eq}\simeq5.64(\Omega_0h^2)eV\simeq2.843\times10^4K$(where we
  set $h=0.71$), $T_0\simeq2.7K$, $a_i=a_{eq}=1$, the scale factor
  at the present epoch $a_0$ would nearly be $1.053\times10^4$,
  then we know at present epoch $N=lna_0\simeq9.262$, we calculate the current value of $\Omega_{\phi_0}\simeq0.705$, decelerating
  factor $q_0\simeq-0.603$ and the transition shift $Z_T\simeq0.672$ in our model, this is consistent with  current observation data.
\section{Summary} \hspace*{15 pt}We have shown that it is possible
to construct a viable cosmological model based on nonlinear
Born-Infeld scalar field. We get the condition to admit a
late-time attractor solution: $u(X_c)>0,u'(X_c)=0$ and $
u''(X_c)<0$ at critical point, this condition is also true for
quintessence model. However, we should emphasized that this
condition is a sufficient condition. The attractive
 behavior of the solution may resolve the "coincidence
 problem" in dark energy models in principle. We investigate the global structure of the
  dynamical system via phase plane analysis and calculate the
  cosmological evolution by numerical analysis with and without the presence of radiation and
  matter, the result shows the universe will evolve to a de Sitter
  like attractor regime in the future and the phantom field energy
  density will dominant whole universe and behave as a
  cosmological constant. An obvious excellent
 character of our model is that the phantom field survive till
 today (to account for the observed late time accelerated
 expansion) without interfering with the nucleosynthesis of the
 standard model(the density parameter $\Omega_{\phi}\simeq10^{-12}$
 at the equipartition epoch), and also avoid the future collapse of
 the universe, it was indicated by Carroll, Hoffman, and Trodden[19]
 \par Theoretical cosmology with phantom models has become
  an active area of theoretical research. However, recently,
some authors showed in [22] that quantum effects(without a
negative energy kinetic term) can also render $\omega<-1$ on
cosmological scales. Since the idea of phantom cosmology is pretty
new, there are many questions remaining open yet.

 \section{Acknowledgement}
 \hspace*{15 pt}This work is partly supported by Shanghai
 Municipal Science and Technology Commission No.04dz05905 and Shanghai Leading Academic Discipline Programme. \\

{\noindent\Large \bf References} \small{
\begin{description}
\item {1.} {S.Perlmutter et.al., Ap.J, 565(1999);\\
             J.L.Tonry et.al., astro-ph/0305008.}
\item {2.} {C.L.Bennett et.al., astro-ph/0302207;\\
             A.Melhiori and L.Mersini, C.J.Odmann, and M.Thodden, astro-ph/0211522;\\
             D.N.Spergel et.al., astro-ph/0302209;\\
             N.W.Halverson et.al., Ap.J.568, 38(2002);\\
             C.B.Netterfield et.al., astro-ph/0104460;\\
             P.de Bernardis et.al., astro-ph/0105296;\\
             A.T.Lee et.al., astro-ph/0104460;\\
             R.Stompor, astro-ph/0105062. }
\item {3.} {R.R.Caldwell, R.Dave and P.J.Steinhardt, Phys.Rev.Lett\textbf{80}, 1582(1998);\\
             B.Ratra and P.j.Peebles, Phys.Rev.D\textbf{37}, 3406(1998);\\
             P,J.Steinhardt,L.Wang,and I.Zlatev, Phys.Rev.Lett.\textbf{82},896(1996);\\
             X.Z.Li,J.G.Hao,and D.J.Liu, Class.Quantum Grav.\textbf{19},6049(2002).}
\item {4.} {C.A.Picon, T.Damour and V.Mukhanov, Phys.Lett.B\textbf{458},209(1999);\\
             T.Chiba, T.Okabe and M.Yamaguchi,Phys.Rev.D\textbf{62}023511(2000);\\
             C.A.Picon, V.Mukhanov and P.J.Steinhardt,Phys.Rev.Lett\textbf{85}4438(2000).}
\item {5.} {J.M.Aguirregabiria,L.P.Chiemento andR.Lazkoz,Phys.Rev.D\textbf{70},023509(2004);\\
             L.P.Chimento,Phys.Rev.D\textbf{769},123517(2004).}
\item {6.} {A.Melchiorri,astro-ph/0406652;\\
             V.Faraoni, Int.J.Mod.Phys.D\textbf{11}, 471(2002);\\
             S.Nojiri and S.D.Odintsov, hep-th/0304131;hep-th/0306212;\\
             E.schulz and M.White, Phys.Rev.D\textbf{64}, 043514(2001);\\
             T.Stachowiak and Szydllowski, hep-th/0307128;\\
             G.W.Gibbons, hep-th/0302199;\\
             A.Feinstein and S.Jhingan, hep-th/0304069.  }
\item {7.} {R.R.Caldwell, Phys.Rev.Lett.B\textbf{545}, 23(2002);\\
             M.Sami,Mod.Phys.Lett.A\textbf{19},1509(2004)\\
             P.Singh, M.Sami, N.Dadhich,Phys.Rev.D\textbf{68},023522(2003)\\
             D.J.Liu and X.Z.Li, Phys.Rev.D\textbf{68}, 067301(2003);\\
             M.P.Dabrowski et.al., Phys.Rev.D\textbf{68},103519(2003);\\
             S.M.Carroll, M.Hoffman, M.Teodden, astro-th/0301273;\\
             Y.S.Piao, R.G.Cai, X.M.Zhang and Y.Z.Zhang, hep-ph/0207143;\\
             J.G.Hao and X.Z.Li, hep-th/0305207;\\
             S.Mukohyama, Phys.Rev.D\textbf{66},024009(2002);\\
             T.Padmanabhan, Phys.Rev.D\textbf{66}, 021301(2002);\\
             M.Sami and T.Padamanabhan, Phys.Rev.D\textbf{67}, 083509(2003);\\
             G.Shiu and I.Wasserman, Phys.Lett.B\textbf{541},6(2002);\\
             L.kofman and A.Linde, hep-th/020512;\\
             H.B.Benaoum, hep-th/0205140;\\
             L.Ishida and S.Uehara, hep-th/0206102;\\
             T.Chiba, astro-ph/0206298;\\
             T.Mehen and B.Wecht, hep-th/0206212;\\
             A.Sen, hep-th/0207105;\\
             N.Moeller and B.Zwiebach, JHEP\textbf{0210}, 034(2002);\\
             J.M.Cline, H.Firouzjahi and P.Martineau. hep-th/0207156;\\
             S.Mukohyama, hep-th/0208094;\\
             P.Mukhopadhyay and A.Sen, hep-th/020814;\\
             T.Okunda and S.Sugimoto, hep-th/0208196;\\
             G.Gibbons, K.Hashimoto and P.Yi, hep-th/0209034;\\
             M.R.Garousi, hep-th/0209068;\\
             B.Chen, M.Li and F.Lin, hep-th/0209222;\\
             J.Luson, hep-th/0209255;\\
             C.kin, H.B.Kim and O.K.Kwon, hep-th/0301142;\\
             J.M.Cline, H.Firouzjahi and P.Muetineau, hep-th/0207156;\\
             G.Felder, L.Kofman and A.Starobinsky, JHEP\textbf{0209}, 026(2002);\\
             S.Mukohyama, hep-th/0208094;\\
             G.A.Diamandis, B.C.Georgalas, N.E.Mavromatos, E.Pantonopoulos, hep-th/0203241;\\
             G.A.Diamandis, B.C.Georgalas, N.E.Mavromatos, E.Pantonopoulos, I.Pappa, hep-th/0107124;\\
             M.C.Bento, O.Bertolami and A.A.Sen, hep-th/020812;\\
             H.Lee et.al., hep-th/0210221;\\
             M.Sami, P.Chingangbam and T.Qureshi, hep-th/0301140;\\
             F.Leblond, A.W.Peet, hep-th/0305059;\\
             J.G.Hao and X.Z.Li, Phys.Rev.D\textbf{66}, 087301(2002);\\
             X.Z.Li and X.H.Zhai, Phys.Rev.D\textbf{67}, 067501(2003).\\
             J.G.Hao and X.Z.Li, Phys.Rev.D\textbf{68},043501(2003). }
\item {8.} {H.Q.Lu, Int.J.Modern.Phys.D\textbf{14},355(2005).}
\item {9.} {R.R.Caldwell et.al., Phys.Rev.Lett.\textbf{91},071301(2003);\\
             G.W.Gibbons, hep-th/0302199;\\B.Mcinnes, hep-th/0112066.}
\item {10.} {S.M.Carroll et.al., Phys.Rev.D \textbf{68},023509(2003);\\
             G.Felder, L.Kofman and A.Starobinsky, hep-th/0208019;\\
             G.W.Gibbons, hep-th/031117; hep-th/0302199;\\
             A.Frolov, L.Kofman and A.Starobinsky, hep-th/0204187;\\
             A.Sen, hep-th/0204143; hep-th/0209122; hep-th/0203211.}
\item {11.} {M.Born and Z.Infeld, Proc.Roy.Soc A\textbf{144}, 425(1934).}
\item {12.} {W.Heisenberg,Z.Phys.\textbf{133},79(1952);\textbf{126},519(1949);\textbf{113},61(1939)
             }
\item {13.} {T.Taniuti,Prog.Theor.Phys.(kyoto) Suppl\textbf{9}, 69(1958).}
\item {14.} {H.P.de Oliveira,J.Math.Phys.\textbf{36}, 2988(1995).}
\item {15.} {H.Q.Lu, T.Harko and K.S.Cheng, Int.J.Modern.Phys.D\textbf{8},625(1999).}
\item {16.} {A.Kamenshchik et.al., Phys.Lett.B\textbf{511}(2001);\\
              N.Bilic et.al., Phys.Lett.B\textbf{535}(2002).}
\item {17.} {J.K.Erickson et al., Phys.Rev.Lett\textbf{88},121301(2002).}
\item {18.} {A.Macorra and H.Vucetich, astro-ph/0212302}
\item {19.} {Sean M.Carroll and M.Hoffman, Phys.Rev.D\textbf{68},023509(2003).}
\item {20.} {P.Singh, M.Sami and N.Dadhich, Phys.Rev.D\textbf{68},023522(2003).}
\item {21.} {E.Witten,Phys.Rev.D\textbf{46}5467(1992),Phys.Rev.D\textbf{47}3405(1993);\\
              K.Li and E.Witten, Phys.Rev.D\textbf{48}853(1993);\\
              D.Kutasov, M.Marino and G.Moore, JHEP\textbf{0010}045(2000);\\
              T.Chiba, astro-ph/0206298.}
\item {22.} {V.K.Onemli and R.P.Woodard,gr-qc/0406098, gr-qc/0204065;\\
             T.Brunier, V.K.Onemli and R.P.Woodard, gr-qc/0408080.}

\end{description}}
\end{document}